\documentclass{iaus}
\usepackage{graphicx}

\title[Ages of PMS Stars] %% give here short title %%
{New Methods for Determining the Ages of PMS Stars}

\author[Naylor et al]   %% give here short author list %%
{Tim Naylor$^1$, N.J. Mayne$^1$, R.D. Jeffries$^2$, S.P. Littlefair$^3$ \& 
Eric S. Saunders$^4$}
\affiliation{
$^1$School of Physics, University of Exeter
Stocker Road, Exeter, EX4 4QL, UK. \\[\affilskip]
$^2$Astrophysics Group, School of Physical and Geographical Sciences,
Keele University, Keele, Staffordshire ST5 5BG UK. \\[\affilskip]
$^3$Department of Physics and Astronomy, University of Sheffield,
Sheffield S3 7RH, UK.\\[\affilskip]
$^4$Las Cumbres Observatory Global Telescope network, 
6740 CortonaDrive, Suite 102, Goleta, CA93117, USA 
}

\pubyear{2008}
\volume{258}  %% insert here IAU Symposium No.
\pagerange{119--126}
\jname{The Ages of Stars}
\editors{A.C. Editor, B.D. Editor \& C.E. Editor, eds.}
\begin{document}

\maketitle

\begin{abstract}
We present three new methods for determining the age of groups of 
pre-main-sequence stars.
The first, creating empirical isochrones allows us to create a robust
age ordering, but not to derive actual ages.
The second, using the width of the gap in colour-magnitude space between
the pre-main-sequence and main-sequence (the radiative convective gap) has
promise as a distance and extinction independent measure of age, but is as
yet uncalibrated.
Finally we discuss $\tau^2$ fitting of the main sequence as the stars 
approach the terminus of the main sequence.
This method suggests that there is a factor two difference between these 
``nuclear'' ages, and more conventional pre-main-sequence contraction ages. 
\end{abstract}

\firstsection % if your document starts with a section,

\section{Introduction}

Good age determinations for pre-main-sequence (PMS) clusters and 
associations are crucial for our understanding of this phase of stellar
evolution.
For example, modelling the interaction of young stars with their 
(presumably) planet forming discs requires observational measurements of
the disc dissipation and stellar spin-up timescales, both
of which require accurate ages for the clusters and associations studied.
Equally, any determination of the mass functions in young groups is
strongly dependent on the assumed age.  
In this contribution we review three methods we are developing for measuring 
the ages of PMS clusters and associations using colour-magnitude diagrams
(CMDs).

\section{Empirical Isochrones} 

\begin{figure}
% \vspace*{-2.0 cm}
\begin{center}
 \includegraphics[width=4.7in]{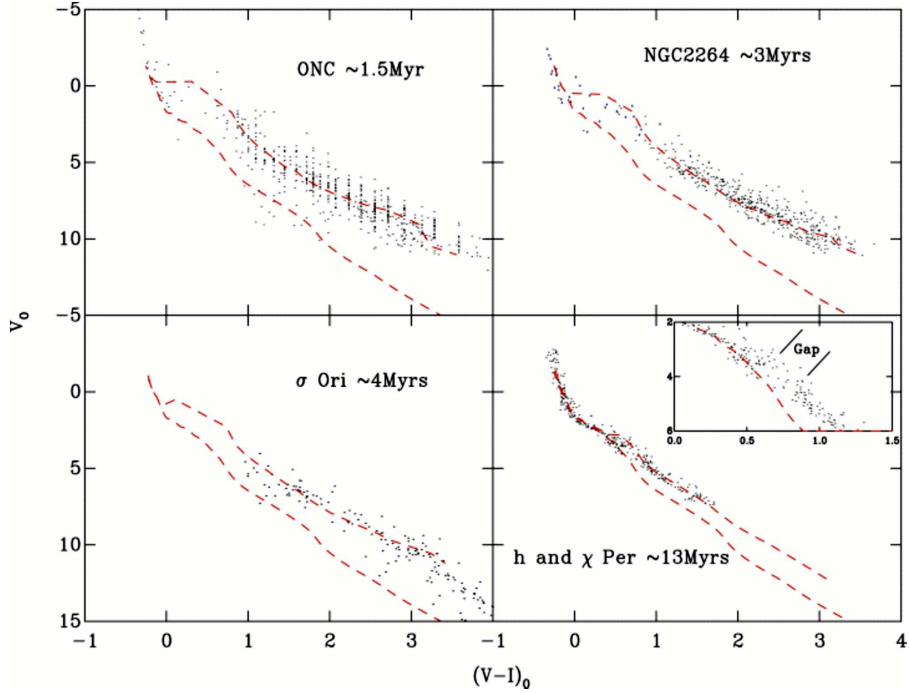} 
% \vspace*{-1.0 cm}
 \caption{
The CMDs for a selection of young groups in absolute
magnitude and intrinsic colour.
In each case the lower red dotted line is the position of the MS, the upper
an appropriate \cite[Siess et al (2000)]{S00} isochrone. 
}
   \label{nf18}
\end{center}
\end{figure}

Figure \ref{nf18} shows CMDs for members of a selection
of young clusters and associations.
The majority of the stars lie on the PMS, which is elevated
in the diagram with respect to the main sequence (MS).
For older groups the stars lie closer to the MS, and this decline in 
luminosity with time is an age indicator.
If the isochronal models for PMS stars were good fits to the data we could 
simply use the best-fitting ages.
However, in practice the data deviate systematically from the isochrone
(e.g. \cite[Bonatto 2004]{B04}; \cite[Pinsonneault et al., 2004]{P04}). 
Furthermore the derived ages can depend on which colour is fitted  
(e.g. \cite[Naylor et al, 2002]{N02}).
An obvious alternative is to abandon attempts to allocate absolute ages,
but develop an age order (or ladder) based on the luminosity of the
PMS.
Simply plotting the data for different groups in the same diagram does not 
lead to useful results because the spread in each sequence is large.
We therefore fit splines through each sequence, and place the splines
in absolute-magnitude, intrinsic-colour space. 
The result of such a procedure is shown in Figure 2 (left), where 
we can see that NGC2264 is older than the ONC, but younger than NGC1960
and about the same age as $\sigma$ Ori.

We used the above method in \cite[Mayne et al (2007)]{M08} to obtain an age 
order for a set of well known clusters and associations, but realised that 
the limiting factor was
the determination of the distance (with which the age is degenerate).
So, in \cite[Mayne \& Naylor (2008)]{MN08} we measured the distances to these 
clusters and associations in a consistent way.
As can be seen in Figure \ref{nf18} the most massive stars are actually on 
the MS.
We therefore fitted these stars to a MS model using $\tau^2$ fitting (see
\cite[Naylor \& Jeffries, 2006]{NJ06} and Section \ref{nuclear}) to derive 
distances.
In Table \ref{tab1} we present the resulting age ordering, including
two more associations from more recent work.
Although we give ages in Table \ref{tab1}, it is worth emphasising that
strictly speaking we only derive an order.
The ages represent an informed average of literature PMS ages for these
groups. 

\begin{table}
  \begin{center}
  \caption{The Empirical Isochrone Age Ladder.}
  \label{tab1}
 {\scriptsize
  \begin{tabular}{|l|c|c|c|c|}\hline 
{\bf Age} & {\bf Groups}  \\ \hline
1Myr    & IC 5146 \\
2Myr    & ONC, NGC 6530  \\
3Myr    & $\lambda$ Ori, $\sigma$ Ori, NGC 2264  \\
4-5Myr  & IC 348, Cep OB3b$^1$, NGC 2362 \\
5-10Myr & $\gamma$ Vel$^2$ \\
10Myr   & NGC 7160 \\
13Myr   & h and \& $\chi$ Per \\
40Myr   & NGC 2547 \\ \hline
  \end{tabular}
  }
 \end{center}
\vspace{1mm}
 \scriptsize{
 {\it Notes:}
  From \cite[Mayne \& Naylor (2008)]{MN08}, except for:
  $^1$Littlefair (in prep);
  $^2$\cite[Jeffries et al (2008)]{J08}
}
\end{table}

\section{Radiative-convective gap}

An interesting feature of Figure \ref{nf18} is the paucity of stars on the
PMS isochrone immediately prior to the point where it joins the MS.
The gap is clear and wide for the youngest groups, but by the age of
h and $\chi$ Per has narrowed almost to the point of invisibility.
The physical explanation for the gap lies in the change of structure 
between the fully convective interiors of PMS stars, and the partially
radiative ones of MS stars.
This drives a change in radius, which happens relatively quickly, leading to
a rapid movement to bluer colours in the CMD, as stars move from the PMS
to the MS.
Hence in \cite[Mayne et al (2007)]{M08} we named this the radiative 
convective gap.

Clearly this change in the size of the gap with age could be used an age 
indicator.
It has two main advantages over PMS (contraction) ages.
First, it uses brighter stars.
Second, since one is measuring a distance, rather than position on
colour-magnitude space, it is independent of errors in distance or
extinction (assuming the latter is uniform). 
Before it can be used as an age indicator, though, it will need to be
calibrated against ages derived from other techniques.

\section{Nuclear (not quite turn-off) ages}
\label{nuclear}

\begin{figure*}
  \centerline{
    \mbox{\includegraphics[width=2.30in]{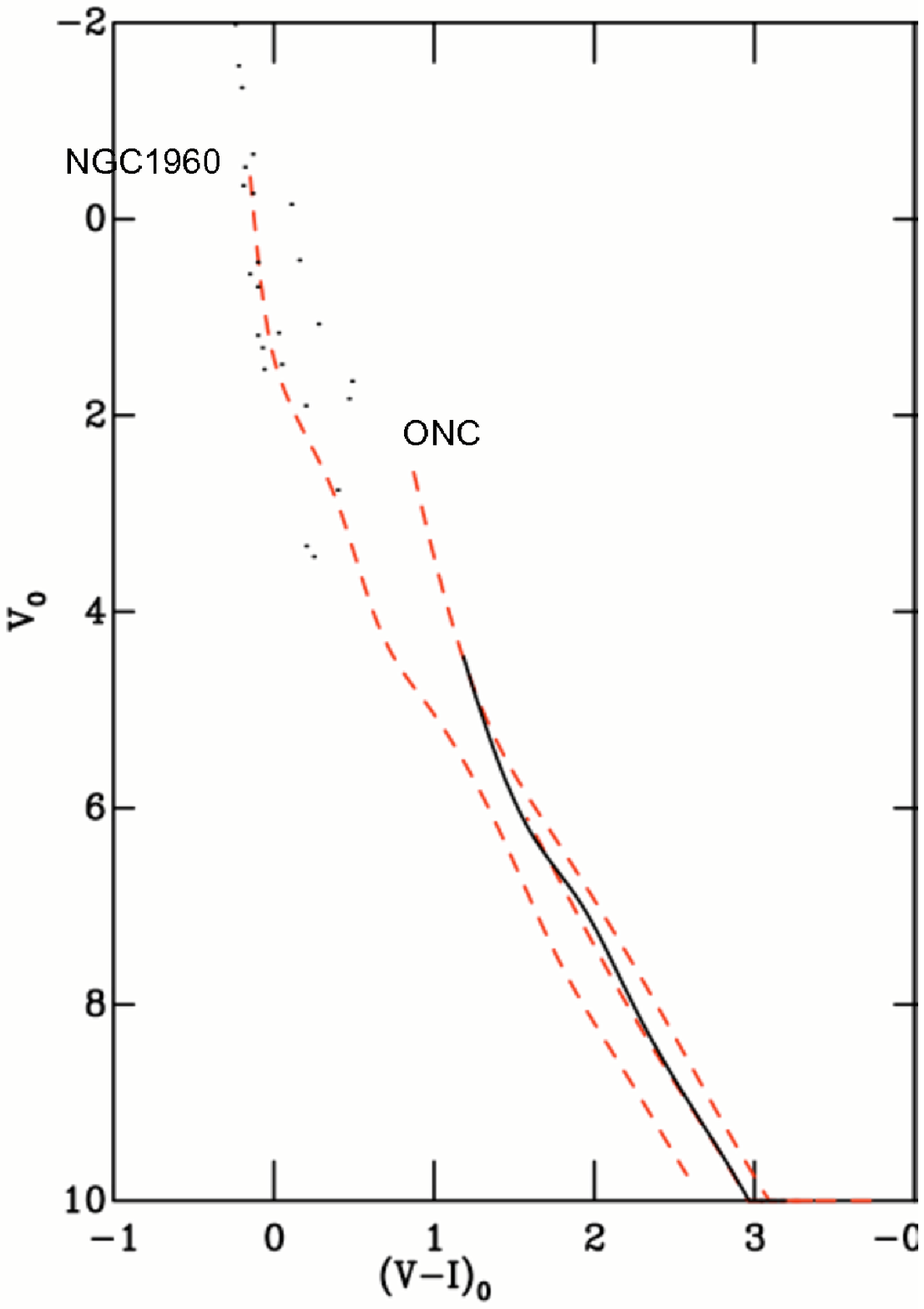}}
    \mbox{\includegraphics[width=2.80in]{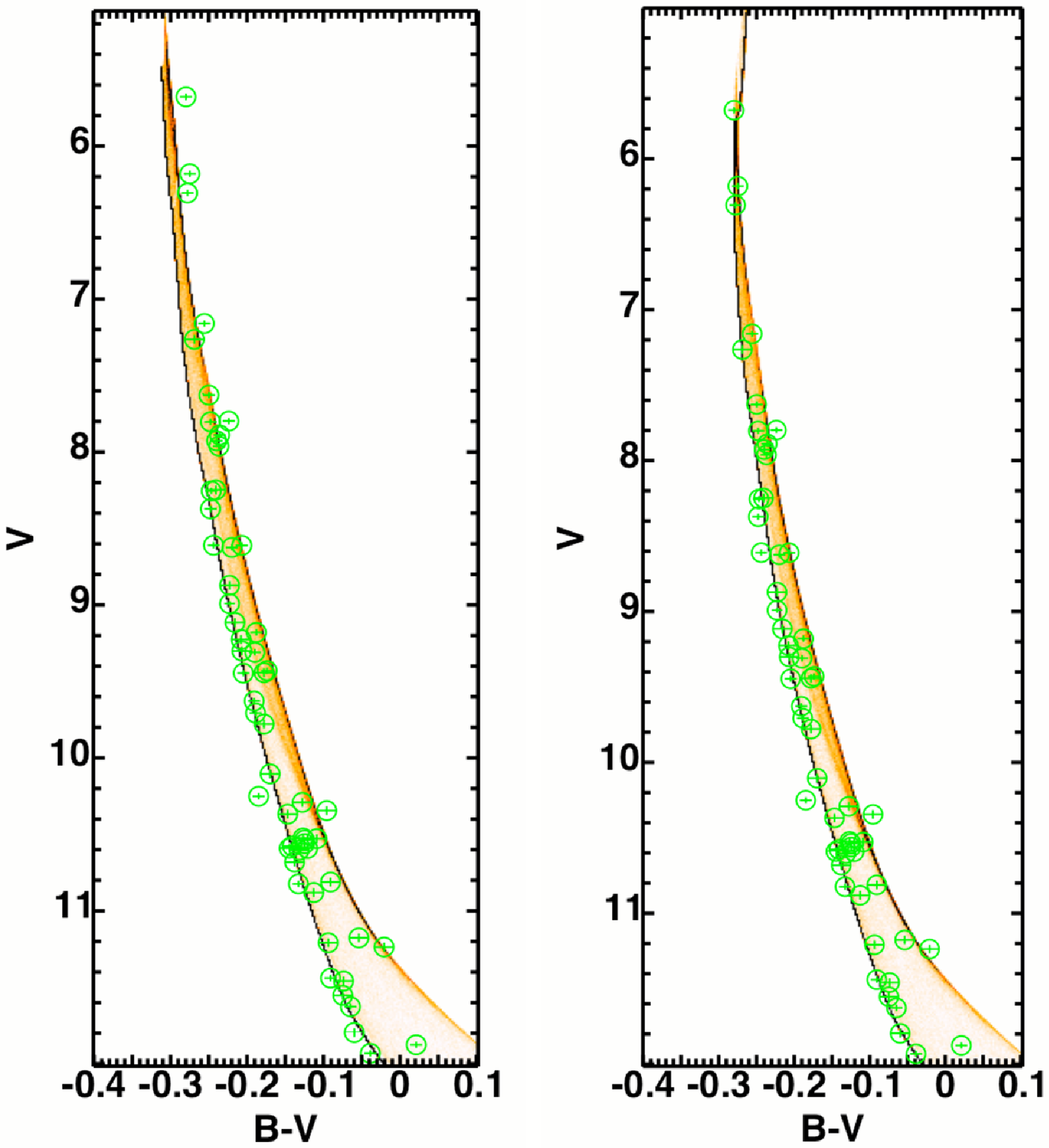}}
\label{both}
  }
  \caption{
{\bf Left}: The empirical isochrone for NGC2264 (black curve), 
compared with other clusters and associations (the red dashed curves). 
The unmarked red dashed isochrone close to that for NGC2264 is $\sigma$ Ori.
{\bf Right}: Geneva-Bessell isochrones fitted to the NGC6530 data of 
\cite[Walker (1957)]{W57}.
Each star has been dereddened using its position in a $U-B$/$B-V$ diagram.
The fit in the left hand panel had the age fixed at 0.25Myr-old, yielding 
$Pr(\tau^2)=0.03$.
On the right the age was a free parameter, found to be 5.5Myr, with 
$Pr(\tau^2)=0.67$.
}
  \label{overView}
  \end{figure*}

Before stars turn off the MS, they evolve redwards away from the 
zero-age MS, driven by their nuclear evolution.
This means that in colour-magnitude space the MS, which normally has a positive
gradient can, near its high-mass terminus, be vertical or even have a 
negative gradient (see Figure 2 (right)).
Although the effect is subtle, if we can fit it, this should provide an age 
indicator.
The best method for doing this is the $\tau^2$ method we described in  
\cite[Naylor \& Jeffries (2006)]{NJ06}, since this allows for the effects of 
binarity, gives reliable uncertainties for the parameters, and provides a 
goodness-of-fit test.
Unfortunately the technique as we described it will not work if the isochrone 
is vertical.
Therefore we first outline the improvements which we have made to the method to 
allow us to address this problem (which will be described in more detail in
Naylor in prep.), before moving on to our results.

\subsection{Improvements to the $\tau^2$ technique}

The technique relies on a finely sampled grid (such as the colour scales of
Figure 2 (right)), which is a model created from of order a million simulated
stars.
We refer to this as $\rho(c,m)$ where $c$ and 
$m$ are the colour and magnitude co-ordinates respectively.
For any given datapoint we calculate $\tau^2$ by multiplying this grid on a
point-by-point basis with a function representing the datapoint and its 
uncertainties (typically a two-dimensional Gaussian).
We represent this as $U_i(c-c_i, m-m_i)$, where $i$ is the index of a datapoint
at co-ordinates $(c_i,m_i)$.
If we now sum the resulting points, and repeat this over all datapoints we 
arrive at the definition of $\tau^2$
\begin{equation}
\tau^2=-2\sum_{i=1,N}{\rm ln}\int{U_i(c-c_i, m-m_i)\rho(c,m){\rm d}c\ {\rm d}m}.
\label{for_tau}
\end{equation}
The best-fitting model corresponds to the lowest value of $\tau^2$.
For example, a simple fit in distance modulus can be viewed as moving the models
up and down in Figure 2 (right) until the ``cross correlation'' 
between the datapoints and the model is maximised.
In practice the simplest way to find the best fitting model is to calculate 
$\tau^2$ for a grid of models covering the range of parameters of interest.

There is a question as to how $\rho$ should be normalised.
In \cite[Naylor \& Jeffries (2006)]{NJ06} we derived a normalisation such that 
Equation \ref{for_tau} reduced to that for $\chi^2$ for fitting a curve to 
data with uncertainties in one dimension.
Unfortunately, when the
isochrone is vertical, this results in an infinity in Equation 17 of that 
paper making it impractical for post-main-sequence fitting.
Instead we now use a normalisation where the integral of $\rho$ between the faintest
and brightest datapoints is one.  
Similarly we demand that the integral of $U$ is one over the entire CMD.

Having found a fit to the data, we must establish whether it is a good fit.  
We do this by calculating the probability that we would obtain our value of
$\tau^2$ from observations, assuming the model was correct.
This is $Pr(\tau^2)$, which we showed how to calculate for no free parameters in 
\cite[Naylor \& Jeffries (2006)]{NJ06}.
Our suggested correction to allow for free parameters, multiplying the values of
$\tau^2$ by $(N-n)/N$ (where $N$ is the number of datapoints and $n$ the number
of free parameters), is not invariant under changes in normalisation.
A better approximation is to subtract the expectation value of $\tau^2$ before
multiplying by $(N-n)/N$, and then add the expectation value on again.

Finally, to find the uncertainties in the parameters we have found a quicker
method than that we presented in \cite[Naylor \& Jeffries (2006)]{NJ06}.
Assuming the minimum value of $\tau^2$ has been found by a grid search, each 
datapoint in the grid has a probability $P$ associated with it 
(via the definition of $\tau^2$) of
\begin{equation}
P=e^{-{\tau^2}/2}.
\end{equation}
By summing the probability below a given $\tau^2$, and dividing by the 
probability summed over the entire grid, one can obtain the probability 
that $\tau^2$ lies below a given value.
This allows one to draw a confidence contour in the parameter space, in an 
identical fashion to that used in $\chi^2$ analysis.

\subsection{Results}

We have fitted UBV photometry for bright stars in NGC6530, NGC2264, $\sigma$ 
Ori, $\lambda$ Ori, NGC2362, Cep OB3b, NGC2547, IC2602 and stars in the 
vicinity of the Orion Nebula Cluster.
For preference we have used data from the 1950s to 1970s of Walker, Johnson 
and collaborators, since this is a relatively homogeneous group of datasets, 
and we find our models fit them well.
We have used the Geneva-Bessell models described in 
\cite[Mayne \& Naylor (2008)]{MN08}.
We first use a $U-B$/$B-V$ diagram to determine the extinction. 
In some cases the extinction is uniform, and we determine its value using 
$\tau^2$ fitting.
In other cases we find the extinction is non-uniform, and we find the 
extinctions on a star-by-star basis by comparison with the isochrone as 
described in \cite[Mayne \& Naylor (2008)]{MN08}.
This is essentially an updated $Q$ method.
We then perform a grid search in both age and distance to derive ages and 
associated uncertainties.
Unsurprisingly our distances are all consistent with those in 
\cite[Mayne \& Naylor (2008)]{MN08}, but for the groups less than 10Myr old, 
we find the ages are a factor 1.5-2.0 larger than those given 
in Table \ref{tab1}.

Before attaching any significance to this result, we questioned whether it could
be due to either our fitting procedure or the models used.
We have experimented with models without convective overshoot, which we find 
give poor fits to the data, and with the Padova models 
(\cite[Girardi et al 2002]{G02}), which we find give similar answers to
the Geneva-Bessell isochrones, though both models have yet to be tested 
exhaustively.
We also find that our age for stars in the vicinity of the ONC (5Myr) is 
similar to that found using the same dataset by 
\cite[Meynet et al (1993)]{MMM93}.
For IC2602 \cite[Mermilliod (1981)]{M81} obtains a nuclear age of 35Myr, 
which compares favourably with our estimate of about 40Myr.
For NGC 2547 we obtain about 45Myr, compared with \cite[Claria (1982)]{C82} 
who obtains 57Myr.
Since our ages are broadly consistent with other turn-off/nuclear ages, we
can rule out some systematic effect from our models or fitting procedure.
We are, therefore, forced to conclude that this is a genuine discrepancy 
between PMS ages and nuclear ages. 

For the groups older than 10Myr it is harder to be definitive about any 
difference.
The problem is we need both 
PMS photometry (to obtain an age on the same scale as 
Table \ref{tab1}) and good photo-electric photometry to obtain a nuclear age.
We can carry out the test for NGC2547 for which we obtain 45Myr, compared with 
PMS and Lithium depletion ages of about 38Myr 
(\cite[Naylor \& Jeffries, 2006]{NJ06}) suggesting the discrepancy decreases 
with age.
For IC2602 the situation is more ambiguous.
The age we obtain (40Myr) is larger than the 
PMS age of 25Myr found by \cite[Stauffer et al (1997)]{S97}, but is 
consistent with the finding of \cite[Jeffries et al (2000)]{J00} that IC2602
is a little older than NGC2547.
Thus the question as to whether this discrepancy disappears for older 
clusters (and if so at what age) must await obtaining further PMS ages
for older clusters.

\subsection{Implications}

It appears we have found a genuine difference between the PMS contraction
and MS nuclear age scales.
It is hard at this stage to decide which scale is correct, but as most modern
work relies on the PMS age scale, it is interesting to examine the
implications of the nuclear age scale being correct.
It would help address two outstanding problems in the area.
First, it is well known that that there appears to be a lack of clusters
in the age range 5-30Myr (\cite[Jeffries et al, 2007]{J07}).
Changing the ages in the way we suggest would fill that gap, especially if
there is a return to the classical age scale for clusters older than 30Myr.
Second there is a problem in the disc-clearing timescale measured from
IR observations of young stars (of order 3Myr, e.g.
\cite[Brice{\~n}o et al, 2007]{B07}) and the 
time required to form a planet by classical core accretion (perhaps 9Myr; 
\cite[Pollack et al, 1996]{P96}).
Whilst there are active attempts to solve this problem by reducing the
theoretical timescale (e.g. \cite[Dodson-Robinson et al, 2008]{DR08} and 
references therein), our work supports 
a different solution - increasing the observed timescale by a factor two.

\begin{figure}
% \vspace*{-2.0 cm}
\begin{center}
 \includegraphics[width=3.5in]{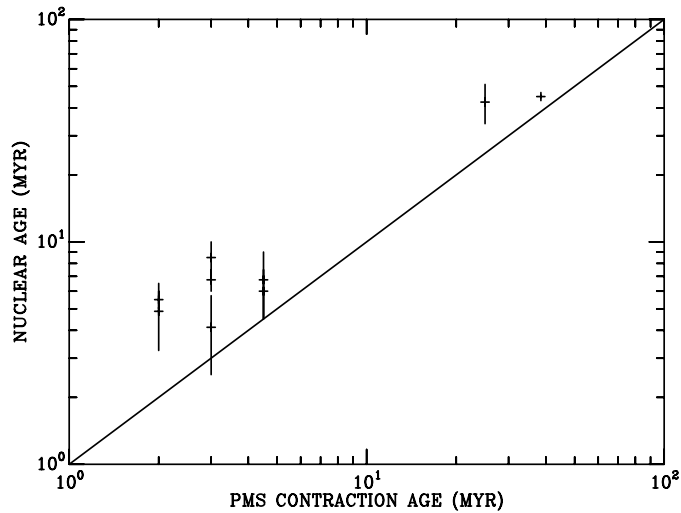} 
% \vspace*{-1.0 cm}
 \caption{PMS vs nuclear ages for young clusters
and associations.
The datapoints would lie on the line if they were equal.
The error bars show the uncertainties derived from the $\tau^2$ fitting.}
   \label{ages}
\end{center}
\end{figure}

%\begin{discussion}

%\end{discussion}

\end{document}